\documentclass{ametsocV6.1}

\title{Event-scale Internal Tide Variability via X-band Marine Radar}

\authors{Alexandra J. Simpson\aff{a}\correspondingauthor{ajoysimpson@gmail.com}, Jacqueline M. McSweeney\aff{b}, James A. Lerczak\aff{c}, Merrick C. Haller\aff{a}}

\affiliation{\aff{a}{School of Civil and Construction Engineering, Oregon State University}, \aff{b}{School of Marine and Atmospheric Sciences, Stony Brook University}, \aff{c}{College of Earth, Ocean, and Atmospheric Sciences, Oregon State University}}

\abstract{A combined radar remote sensing and in situ data set is used to track packets of nonlinear internal waves as they propagate and shoal across the inner shelf ($\sim$40 $m$ - 9 $m$). The dataset consists of high space-time resolution (5$m$, 2$min$) radar image time series collected over a 10 $km$ radial footprint, with over a dozen synchronous and co-located moorings measuring temperature, salinity, and velocity throughout the water column. The internal bores and higher-frequency internal waves that make up the internal tide are tracked in the radar image time series and, therefore, provide continuous cross-shore speed and angle estimates as the waves propagate across the inner shelf. We compare radar-estimated speeds to those estimated from moorings and confirm a cross-shore shoaling profile that deviates from linear theory. We additionally use combined remote sensing and in situ data to perform a detailed analysis on four consecutive internal tides. These analyses reveal intra-packet speed variability, tide-tide influences, reflected internal waves, and an instance of internal wave polarity reversal observed in the radar and moorings.}

\begin{document}

\maketitle

%
%
%
%
%
%

%




\section{Introduction}
The inner shelf is a dynamically rich region of the ocean characterized by nonlinear processes that bridge the shoreline, through the surf zone, to the shelf's edge. Nonlinear internal waves (NLIWs) are a predominant feature of the inner shelf and have strong horizontal (O(1 $m/s$)) and vertical (O(0.1 $m/s$)) velocities that drive large vertical displacements (O(10s $m$)) of the pycnocline. NLIW implications can include the generation of intense nearshore mixing and the transport of nutrients and pollutants \citep{Kumar2021}. Additionally, the presence of NLIWs can strongly influence acoustic propagation and the operation of sonar systems, which may inhibit navigation, object detection, and communication with other vessels. While there is a large body of research that includes modeling, laboratory measurements, and field observations describing internal waves in deep water, less is known about NLIW shoaling and transformation as the waves propagate through the shallow ($<$50 $m$) coastal waters of the inner shelf.

In deep water, NLIWs exist as waves of depression,  meaning that their amplitude is a downward depression of the pycnocline. As they propagate shoreward into shallower waters, they undergo nonlinear processes that may influence their speed and shape characteristics. For instance, if not fully dissipated during the shoaling process due to depth-induced breaking, the NLIWs may undergo a polarity reversal from waves of depression to waves of elevation (e.g., \citet{orr2003nonlinear,scotti2008shoaling,shroyer2009observations}). This wave transformation occurs as a NLIW of depression reaches a pycnocline that is at or below the mid water column (i.e., the critical depth). In shelf waters, the relative layer depths change due to changing total water depth. Generally, the wave of depression forms a dispersive tail containing waves of elevation and waves of depression. The lead wave of depression is eventually destroyed, thus leaving a lead wave of elevation in its place \citep{shroyer2009observations}. This phenomenon has been documented in both laboratory \citep{helfrich2006long} and field observations \citep{orr2003nonlinear, shroyer2009observations}. In addition, numerous models have been derived to describe NLIW shoaling. Weakly nonlinear equations include the Korteweg-de Vries (KdV) and extended KdV (eKdV) \citep{lee1974generation,djordjevic1978fission, choi1999fully}. Fully nonlinear solutions are also possible through the Dubriel-Jacotin-Long (DJL) equation \citet{long1953some}. NLIW speed is a balance between nonlinear and dispersive processes, and evolves with shoaling water depths.

Traditional in situ methods for observing NLIWs utilize thermistor chains and ADCPs (e.g., \citep{orr2003nonlinear,colosi2018statistics}), turbulence profilers (e.g. \citep{becherer2020turbulence}), and/or acoustics (e.g. \citep{moum2006pressure}). Vertically-profiling in situ sensors can also provide a local characterization of NLIW shapes at the point of deployment. In addition, by using triangulation of NLIW arrival times from several sensors, wave speed and angle can be determined \citet{alford2010speed,mcsweeney2020observations}. However, these speeds and angles represent averages over the triangulation array and these arrays are spaced at relatively low spatial resolution due to the constraints and sparsity of mooring deployments. Therefore, high resolution cross- and alongshore variability is difficult to capture using point measurements alone.

Alternately, an advantage of remote sensors is that they offer a synoptic picture of the region of interest, essentially providing thousands of point observations over regions of several kilometers. Synthetic Aperature Radar (SAR) on satellites is a commonly used  sensing tool for internal waves because of its large spatial footprint (O(100$km$)) at relatively high spatial resolution (O(10$m$)) (e.g. \citet{apel1983nonlinear,gasparovic1988overview, da1998role,liu2014tracking}). NLIWs are imaged by radar due to their associated surface currents. Generally, where NLIWs induce convergent surface currents there is an increase in surface roughness at short length scales, leading to heightened radar backscatter. Conversely, in regions of surface divergence, the relatively smoother water surface returns very little backscatter. These influences on backscatter intensity yield images of NLIWs that contain alternating bright and dark bands. Such remotely sensed observations provide snapshots of spatial complexity capable of revealing inter-packet dynamics. For example, \citet{hsu2000study} observed phase shifts due to wave-wave interactions in snapshots from satellite SAR, and \citet{magalhaes2021using} used SAR to look at Mach-stem wave-wave interactions. However, due to orbital constraints, most satellite sensors have a relatively sparse sampling period for image collection (hours to days), thus making spatial evolution difficult to track. On the other hand, terrestrial remote sensors can sample more continuously. 

As an alternative to SAR, marine radars (X-band) have a long history of being utilized for NLIW imaging \citep{alpers1985theory,chang2008composite,ramos2009determination,celona2021automated}. Marine radars can sample over spatial scales of several kilometers with resolution on the order of several meters, and are deployed on vessels or shore-based towers. A real utility of marine radars is that they can image internal waves with high temporal resolution (O(1 min)) to allow tracking in time, which provides the capability of high resolution NLIW speed data and spatial evolution (e.g. \citet{mcsweeney2020alongshore,celona2021automated}).
 Recent work using marine radar has utilized the Radon Transform for NLIW tracking, with much higher cross- and along-shore spatial information than can be obtained with in situ measurements, but only focused on the lead bore \citep{celona2021automated}, or several homogeneously oriented waves \citep{ramos2009determination}. The complex NLIW packet dynamics, such as wave merging, wave-wave interactions, and dispersion remains relatively unexamined via remote sensing.

In this paper, we concentrate on marine radar (X-band) observations collected over two months on the south-central California coast during the fall of 2017. Using space-time diagrams that demonstrate complex intra-tide NLIW behavior, and focus on event-scale NLIW variability that has not been depicted by previous in situ and remote sensing studies. In this work we will refer to packets of NLIWs generated on tidal time scales as "internal tides", the first or leading wave of an internal tide as the "bore", and the remaining non-leading NLIWs in an internal tide as trailing waves. 

Herein, we first provide a site description and an overview of the sampling strategy. Then we detail the developed tracking scheme for estimating cross-shore phase speed and angle of NLIWs, which is effective for both the large amplitude leading bore of an internal tide, as well as the higher frequency trailing waves that follow and make up the rest of the wave packet. Observations are compared to measurements from an array of moorings and confirms that the cross-shore speeds deviate from linear theory. Using the space-time diagram approach, we have revealed instances of wave direction reversal at shallow depths indicating evidence of bore reflection. Additionally, three nearly-consecutive tides are examined in close detail, revealing evidence of alongshore variability, intra-packet speed variability (internal wave dispersion), and an instance of polarity reversal.

\section{Study site \& observations}

\begin{figure}[h]
\centerline{\includegraphics[width = 30pc]{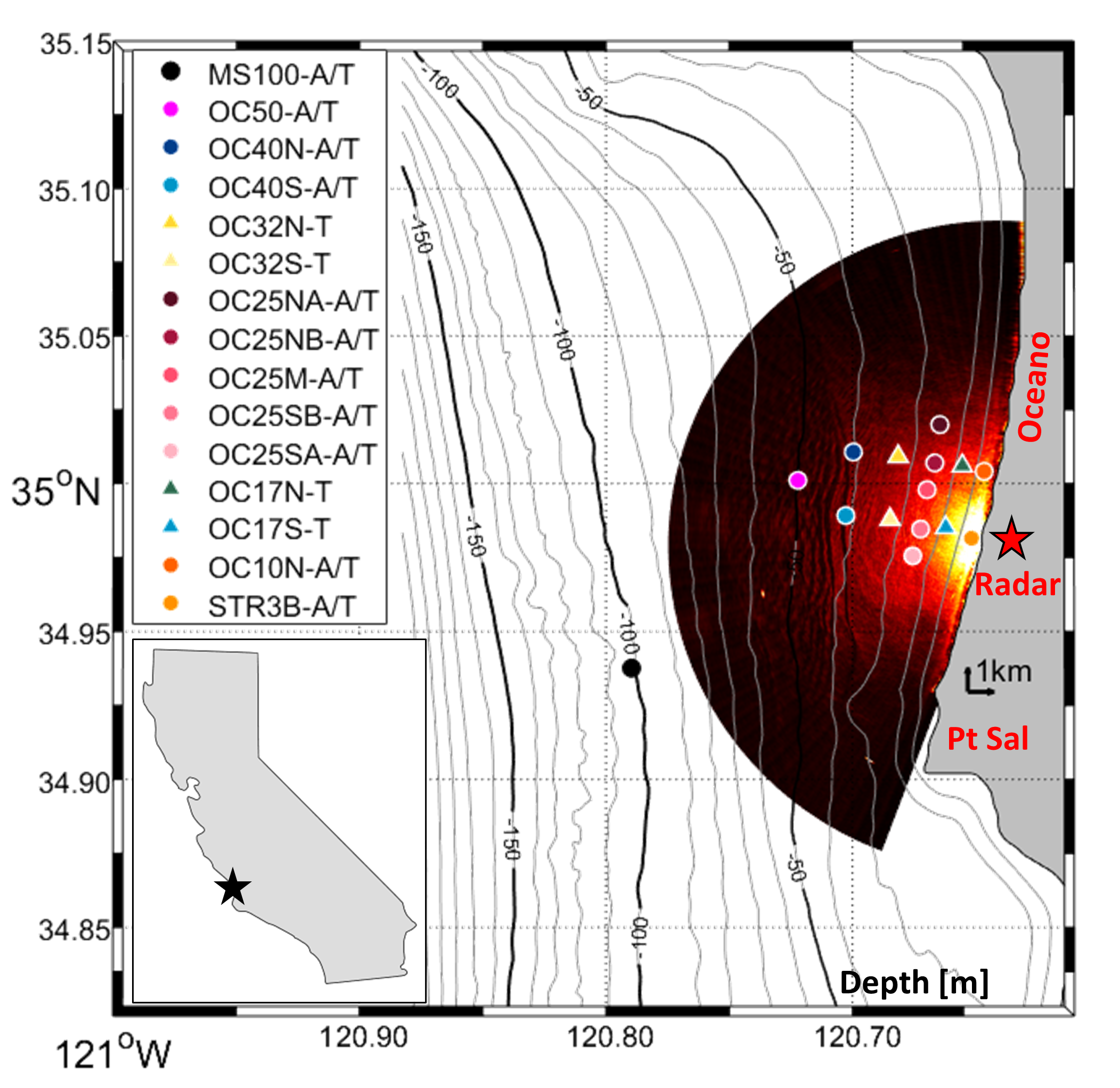}}
\caption{Example image from shore-based radar, with mooring locations indicated. Circles indicate sites with both full water-column temperature and current velocity data, whereas triangles note sites with only the temperature data. This study site, which is just north of Pt Sal is indicated as a black star in the lower left map of California. The location of the radar is indicated with a red star onshore of the mooring array. The legend lists the names of the moorings, where the number represents water depth. The ``A'' in the mooring names indicates an upward-looking ADCP and ``T'' indicates a string of temperature sensors.}
\label{fig: region}
\end{figure}

This study utilizes data collected during the Inner Shelf Dynamics Experiment (ISDE) that is further described in
\cite{lerczak2019untangling,Kumar2021}. The ISDE was a multi-institution field experiment conducted from September-November 2017 near Point Sal, California. The focus of the ISDE was the transition zone between midshelf waters (O(100 $m$)) and the surf zone, and utilized a broad range of moored, towed, land-based, and airborne instrumentation. In this work, we focus on measurements of internal waves from a tower-mounted, X-band marine radar, as well as a subset of the in situ mooring data from the region offshore of Guadalupe, California, termed the "Oceano array". The radar footprint and locations of the moorings are shown in Figure \ref{fig: region}. 

The radar observations were collected using a commercial Koden X-Band marine radar mounted atop a 100-ft tower. The tower was located at the Chevron Oil restoration site in Guadalupe, California, directly onshore of the Oceano Mooring array as indicated in Figure \ref{fig: region}. Radar backscatter intensity images were collected over a footprint of 12 $km$ radius with a sampling period of approximately 1.2 $sec$. These raw images were then averaged over 64 rotations ($\sim$1.3 $min$) to eliminate signatures of surface gravity waves and instead enhance the imaging of the lower frequency hydrodynamic features. Internal waves were frequently imaged from the edge of the domain 12 $km$ offshore (50 $m$ water depth), and tracked as they propagated to shore towards the radar. Given the high resolution in space ($O(1 m)$) and time ($O(1 min)$) of the radar imagery, this data set provided the unique ability to estimate high-resolution space and time properties of the internal wave packets as the waves propagate across the shelf, with synchronous mooring data. For additional specifications of the radar deployment, the reader is referred to \citep{haller2019radar}.

The spatio-temporal overlap of the radar footprint and the Oceano mooring array provided the opportunity to directly compare remotely sensed and directly measured in situ observations of NLIWs. The moorings provided corroboration of bore arrivals, observations of NLIW shape, estimation of background stratification, and sparse estimations of NLIW speeds at the mooring locations \citep{mcsweeney2020observations}. We utilized data from an array of 15 moorings deployed in 100$m$ to 9$m$ depth, which included both a string of temperature sensors and an upward-looking ADCP. We focused primarily on the Southern cross-shore axis of the mooring array (OC40S, OC32S, OC25SB, OC17S, STR3B) because the radar imaging tended to be better in this region of the domain due to winds being more frequently from the southwest. Further description of the mooring instrumentation and data processing can be found in \cite{mcsweeney2020observations}. The temperature and velocity data provided in situ knowledge of NLIW shape. In addition, we utilized bore arrival times determined using a semi-automated detection of filtered pycnocline displacements \citep{mcsweeney2020observations}. Generally, two internal bores were detected within a semidiurnal tidal period (i.e., roughly every 6 hours). Arrival times at the mooring sites were used to identify those specific bores in the radar data. This is an important component of our analysis, as bore identification in radar imagery is  not always obvious. For instance, the leading bore is not always the first or brightest wave imaged in an internal tide. We additionally utilize the mooring-estimated bore speeds as comparison data. Those speeds were determined by \citep{mcsweeney2020observations} using a triangulation method of arrival times as described in \citet{thomas2016horizontal}. In this work, stratification data from the moorings are also used to estimate theoretical linear internal wave phase speeds, as described in Section \ref{sec:methods}.\ref{sec:methods:insitu}.

\section{Methods}
\label{sec:methods}
We use X-Band radar imagery to estimate the continuous cross-shore speed and angle of NLIWs as they propagated across the inner shelf. This section details the developed image processing technique for internal bore and high frequency solitary wave tracking and outlines the use of mooring observations in conjunction with the remote sensing observations. The stratification measured by the moorings is used to to calculate theoretical linear wave speed, and to look in detail at wave shape. 

\subsection{X-Band radar}
To estimate NLIW speed and angle from the radar, we utilize space-time (aka Hovmöller) diagrams \citep{hovmoller1949trough} where intensity is extracted along a spatial transect through time. These diagrams are a commonly used technique throughout coastal oceanography for analysis of remote sensing data (e.g. \citep{holman1993application,chickadel2003optical,catalan2014microwave}), and have been previously used to image internal wave propagation \citep{ramos2009determination}. Herein, we use the space-time diagrams to track NLIWs and estimate propagation speed and angle through phase relationships between nearby transects. Multiple NLIWs are traced throughout an internal tide and individual waves in the space-time diagrams are linked to in situ observations to reveal their subsurface profiles. 

\begin{figure}[h]
\centerline{\includegraphics[width=37pc]{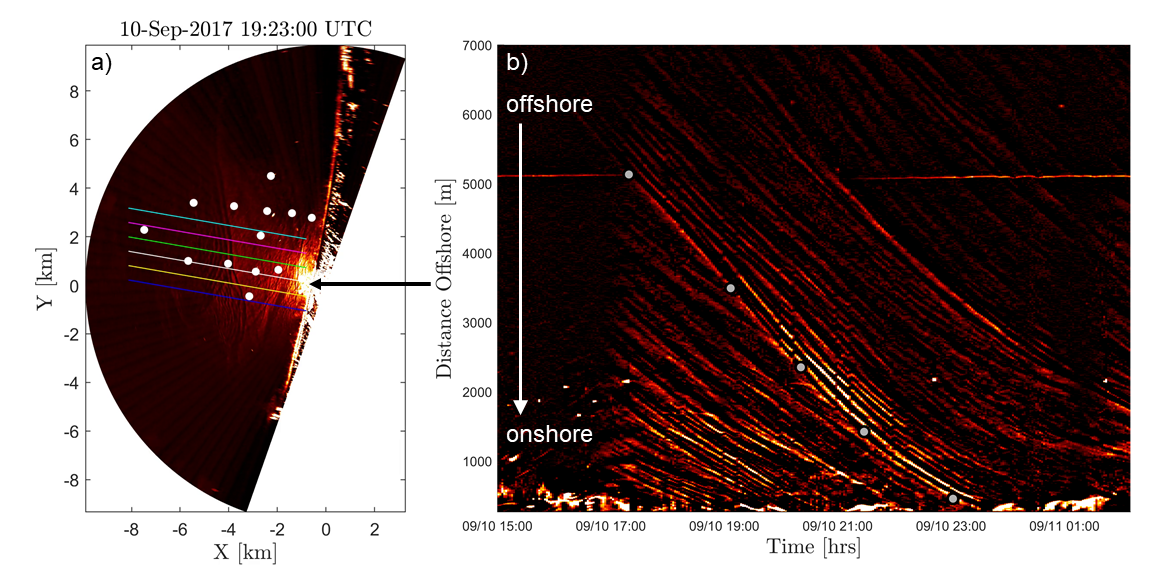}}
\caption{a) Sample radar image; internal waves are visible as bright and dark bands parallel to shore. Mooring locations are given as white dots. Six colored lines are transects used for space-time diagram extraction. b) Space-time diagram extracted from the white transect in panel a. Arrival times at the moorings overlapping the white transect are shown as grey dots, aligned with the lead internal bore.}
\label{fig: radar plus timestack}
\end{figure}

To generate the space-time diagrams that can be compared directly to mooring measurements, six cross-shelf transects aligned with the axes of the mooring array are used for speed and angle estimations (Figure \ref{fig: radar plus timestack}a). The transect that crosses the southern arm of the mooring array (white in Figure \ref{fig: radar plus timestack}a) is the primary transect used for analysis due to both its close proximity to moorings OC40S, OC32S, OC25SB, OC17S, and STR3B and the fact that  radar imaging was better in the southern region of the radar footprint due to predominantly southwesterly winds. While the transects are roughly aligned with the wave propagation direction, the slope of features in the space-time diagrams represent the component of speed in the transect direction, rather than the true vector speed. Therefore, the estimated speeds in transect coordinates are corrected using wave angle to find the true wave speed. Wave angle is found using the phase relationship of the lead bore between the six transects. 

Figure \ref{fig: radar plus timestack}b shows a sample space-time diagram extracted along the white transect (third from the bottom in panel a), with distance offshore on the vertical axis and time on the horizontal axis. The bright, curvilinear streaks are the imaged NLIWs propagating along the transect through time. In order to normalize the range-dependent roll-off of backscatter intensity that occurs with marine radars, the space-time diagram is lowpass filtered from the raw extracted background intensity using a forward-backward filter as described in \citet{gustafsson1996determining} to achieve a zero-phase digital filter in the along-transect dimension with a window size of 400$m$. Next, this range-filtered space-time matrix is subtracted from the original matrix, yielding the range-normalized space-time diagram used for analysis. 

The orientation of the space-time diagram is such that the waves can be considered to propagate from the upper left corner to the lower right corner of the diagram. One can initially detect that the slope of the waves decreases as the waves approach shore, indicating a slowing in speed as the waves shoal. The lead bore of this packet of internal waves is identified by grey dots in Figure \ref{fig: radar plus timestack}b. These represent the arrival times of the lead bore at the moorings in the southern axis of the array, which will be discussed in more detail in the next subsection. By using the arrival times of the lead bore in the moorings, we are able to determine with certainty which streak in the space-time diagram is the lead bore, which can be difficult to determine using the radar data alone. 

\begin{figure}[h]
\centerline{\includegraphics[width=37pc]{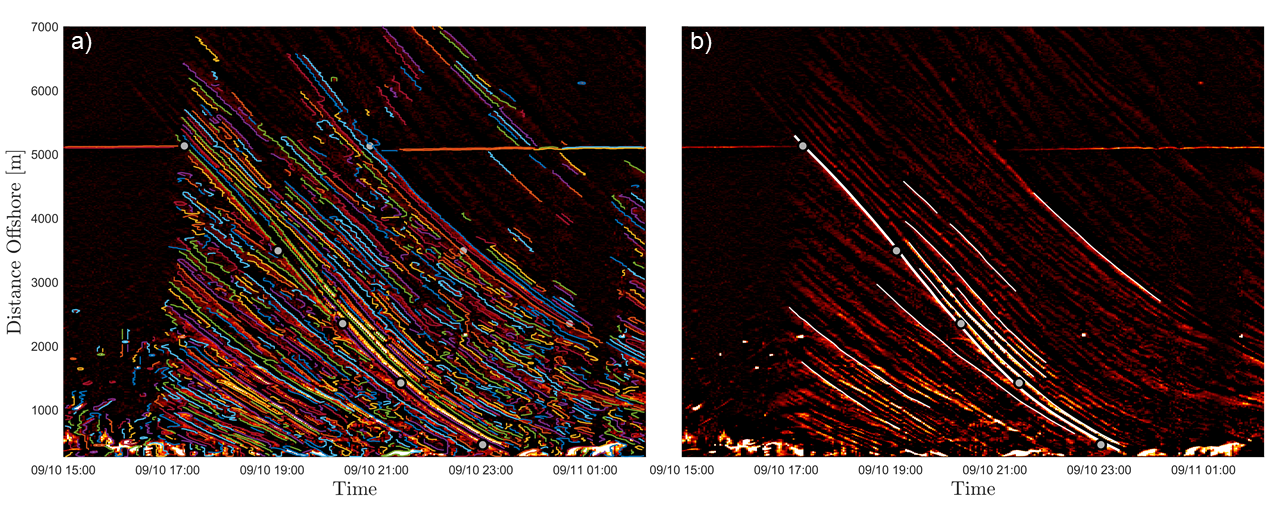}}
\caption{a) Space-time diagram showing edge link tool for identifying edges of waves in the image. b) Space-time diagram with multiple traced internal waves. In both panels, grey dots indicate arrival times of the lead bore in the Oceano mooring data.}
\label{fig: space-time diagram methods}
\end{figure}

For the tracking of individual waves in the space-time diagram, an edge detection and identification scheme was devised. First, a standard Canny edge detector \citep{canny1986computational} was applied to the image. While the edge detection alone is effective in providing locations of sharp intensity gradients, these gradients are not always continuous in coordinate space. Thus, an edge linking tool \citep{KovesiMATLABCode} was used to link coordinates of consecutive edge points in order to form continuous curvilinear features. Each coordinate group corresponded to a different curvilinear feature in the space-time diagram. The edge groups are demonstrated as colored lines in Figure \ref{fig: space-time diagram methods}a.  For consistently imaged waves, a single group encompassed the entire length of the wave; for features with imaging gaps, the wave was represented by several (typically less than three) edge groups. From the linked edge coordinates, we manually determined which edge group corresponded with which wave. Once all edge groups corresponding to a single wave were identified, a locally weighted regression was fitted to the edge group indices to create a smooth trace of the wave. The lead bore is traced with a bold white line in Figure \ref{fig: space-time diagram methods}, Many of the high frequency waves preceding and following the lead bore were also traced. 

Traces of waves in the space-time diagram are in coordinates of range along transect, $r$, and time, $t$. From these values, the slope of the traces are estimated, yielding a value for $c_t$, or speed in the transect coordinates. $c_t$ is estimated using a finite difference scheme: 
\begin{equation}
    c_{t_i} = \frac{r_{i+1} - r_{i}}{t_{i+1}-t_i}
\end{equation}

The lead bore was traced in all six transects. By using the phase relationship of the detected bore among groups of transects, wave angle, $\theta_w$, was determined. Specifically, to correct speed along the white transect, the transect just to the north (green) and just to the south (yellow) are used. Actual wave speed is then found, $c_w = c_t\cos(\theta_w)$. Estimated values of $\theta_w$ ranged from 265 to 290 degrees (compass angle waves are coming from). These angles yield up to a 15 degree offset from the transect direction, or up to 3 percent adjustment in speed values, which is small though still utilized in the speed profiles for best results. 

\begin{figure}[h]
\centerline{\includegraphics[width=37pc]{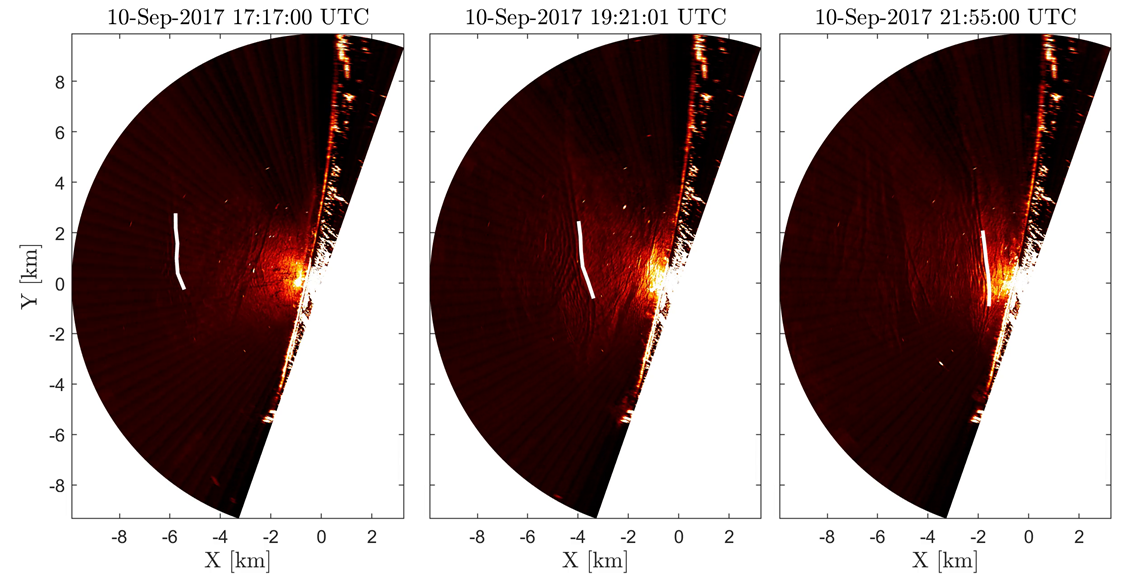}}
\caption{Example of three radar frames with the lead internal bore traced in white, as detected from a series of six space-time diagrams.}
\label{fig: three radar frames}
\end{figure}

With knowledge of the lead bore location in time along each transect, it is also possible to link the wave among the transects, and trace the wave in the original radar images. Figure \ref{fig: three radar frames} shows the identified lead bore in spatial coordinates over three time stamps. This demonstrates some along-shore complexity in the bore's shape.  

\subsection{In situ}
\label{sec:methods:insitu}
Temperature, salinity, and velocity profiles measured by the moorings reveal subsurface NLIW shape, as well as the background stratification that serves as the wave guide through which the NLIWs propagate. To estimate theoretical linear internal wave phase speeds, the background stratification was estimated at each mooring location by computing a sorted density product, which follows \cite{winters1995available}. The sorted product provided a representation of the stratification over a 6 hour window without requiring averaging over the packet of internal waves, which could potentially bias the stratification. Using the sorted density, we calculated linear phase speed, $C_0$, from the eigenvalue problem

\begin{equation}
    \frac{d^2\Phi}{dz^2}+\frac{1}{C_0}(N(z)^2-\omega^2)\Phi=0
    \label{eqn: linear speed}
\end{equation}

\noindent where $\Phi(z)$ is the vertical density structure and a solution to the eigenvalue problem, $\omega$ is the frequency taken as 12.42 hr, and $N(z)$ is the Brunt-V\"{a}is\"{a}l\"{a}, or buoyancy frequency, computed from the sorted density. The eigenvalues $\frac{1}{C_0}$ were found, and only the mode one phase speeds were considered. 

\begin{figure}[h]
\centerline{\includegraphics[width=40pc]{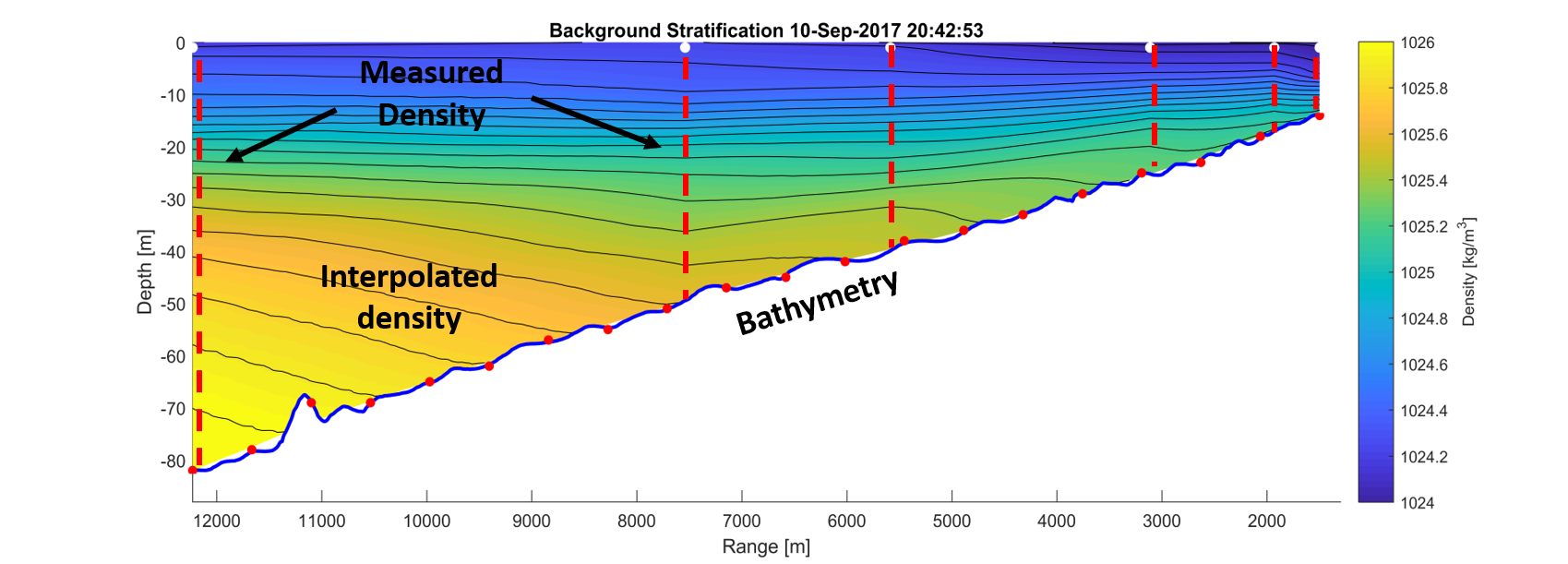}}
\caption{Example of continuous cross-shore stratification found through sigma interpolation of the sorted density product in the cross-shore direction between the discrete locations of the moorings.}
\label{fig: interpolated density}
\end{figure}

The radar-estimated speeds were computed as a continuous cross-shore profile; however, the density used for computing $C_0$ was only measured at the mooring locations, or five discrete points along the cross-shore transect. Thus, to compute a continuous cross-shore linear speed from the mooring observations, we used a sigma-coordinate interpolation method to find a background stratification along the transect. The sorted density profiles from the five moorings were first scaled by the total depth at each mooring location. A linear interpolation was then performed in the cross-shore between the moorings (discretized to 10$m$), before re-scaling back to the original depths. An example of this interpolated density product is shown in Figure \ref{fig: interpolated density}. A value for $C_0$ can then be computed from significantly more density profiles throughout the transect. For our purposes we estimated theoretical linear wave speed using density profiles at the red dots indicated along the bathymetry profile - considerably more discrete locations than at the measured density profiles alone.   

\section{Results}

\subsection{Cross-shore shoaling speeds of internal bores} 

Using the described detection routine, we estimated the speeds and angles of the lead bores for a total of seven internal tides between September 9-17. To choose the bores used for analysis, we first generated space-time diagrams along each of the six transects for every internal tide detected by the moorings throughout the September-October experiment. The time frames for the space-time diagram extraction were chosen by using start times an hour before bore arrivals at OC40S and end times an hour after bore arrivals at STR3B. We then chose the space-time diagrams for analysis that had both 1) consistent imaging of NLIWs in the space-time diagram from offshore to onshore (Figure \ref{fig: radar plus timestack}) during the extracted time window, and 2) bore arrivals detected at all of the transect-overlapping moorings (OC40S, OC32S, OC25S, OC17S, STR3B). This led to seven total bores and their respective internal tides. Each of the seven chosen tides were processed for lead bore speed and angle, where the angle was used to correct transect-direction wave speeds to actual wave speeds.

\begin{figure}[h]
\centerline{\includegraphics[width=40pc]{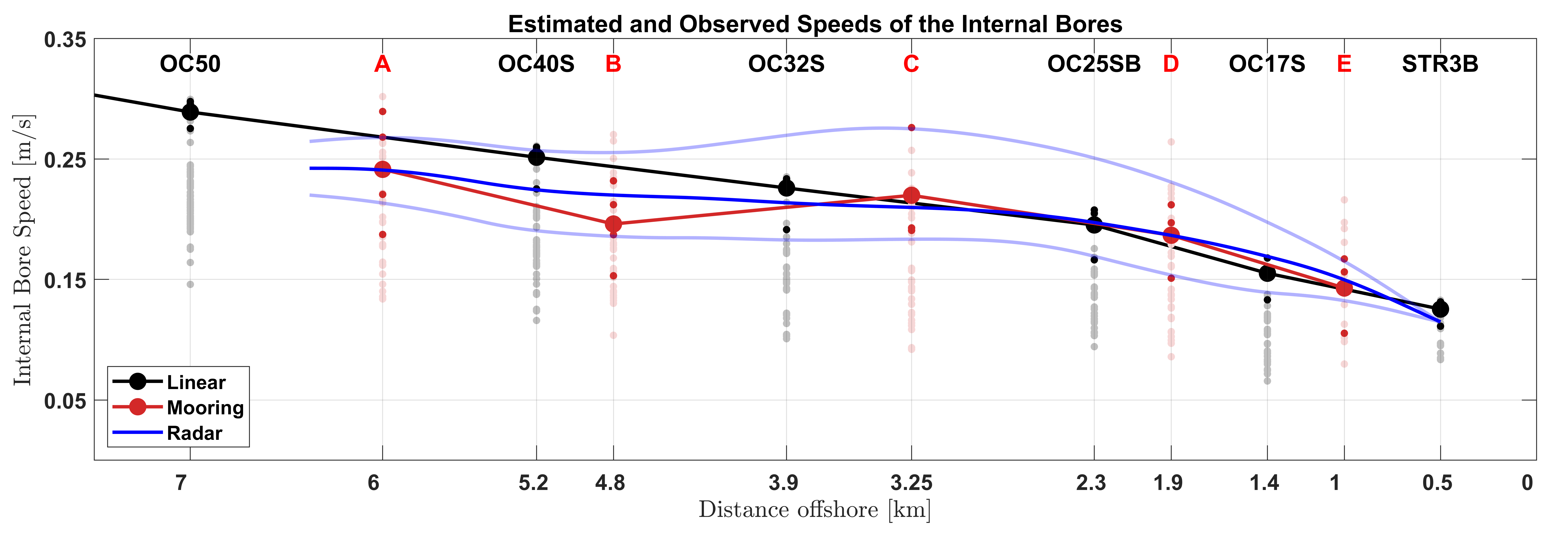}}
\caption{Summary of lead bore speeds estimated from the radar space-time diagram (blue) compared to mooring-estimated speeds (red) and linear theory (black). The ranges of speeds for each source are shown in lighter colors.}
\label{fig: compare mcsweeney}
\end{figure}

Figure \ref{fig: compare mcsweeney} is adapted from \citet{mcsweeney2020observations}, with radar-estimated speeds added to the figure. The locations of the moorings are shown as black dots and labeled using their respective names (see Figure \ref{fig: region}). Linear speed estimations (Equation \ref{eqn: linear speed}) are provided at the mooring locations using the density profiles at the time of the time of the bore arrivals. Linear speeds for all of the detected bores are shown as small grey dots. Linear speed speeds for only the analyzed seven bores are plotted as small black dots, and the mean of those seven shown as large black dots connected in the cross-shore via a black line.
The mooring-estimated speeds using bore arrival times are estimated at locations between the moorings via the triangulation method and indicated by red letters A-E. All of the mooring-estimated speeds are shown as small faded red dots, and the speeds for the seven chosen bores are shown as small solid red dots. The mean mooring speeds are shown as larger solid red dots and connected via a solid red line.  The cross-shore speeds of the seven radar-analyzed bores are averaged and plotted as a solid blue line. The max and min speeds of the seven bores is plotted as faded blue lines. 

The cross-shore trend in radar-estimated speeds is very close to the identified trend in mooring-estimated speeds. It can also be seen that the range of radar-estimated speeds nearly encapsulates the full range of mooring-estimated speeds for the seven analyzed bores. However, the linear estimated speeds do not ever reach the maximum observed speeds, rather maintain a nearly linear profile in the cross-shore. As discussed by \citet{mcsweeney2020observations}, the observed bore speeds are slower than linear theory offshore of 32 $m$ water depth, and faster than linear theory inside of the 32 $m$ depth. Additionally, linear theory predicts that the bores will be slowing consistently from 50 $m$ inshore. However, in the mooring and radar observations, the bores maintain their offshore speed until they reach water depths between 25-17 $m$ before beginning to slow. 

\subsection{Summary of Internal Bore Speeds}

\begin{figure}[h]
\centerline{\includegraphics[width=42pc]{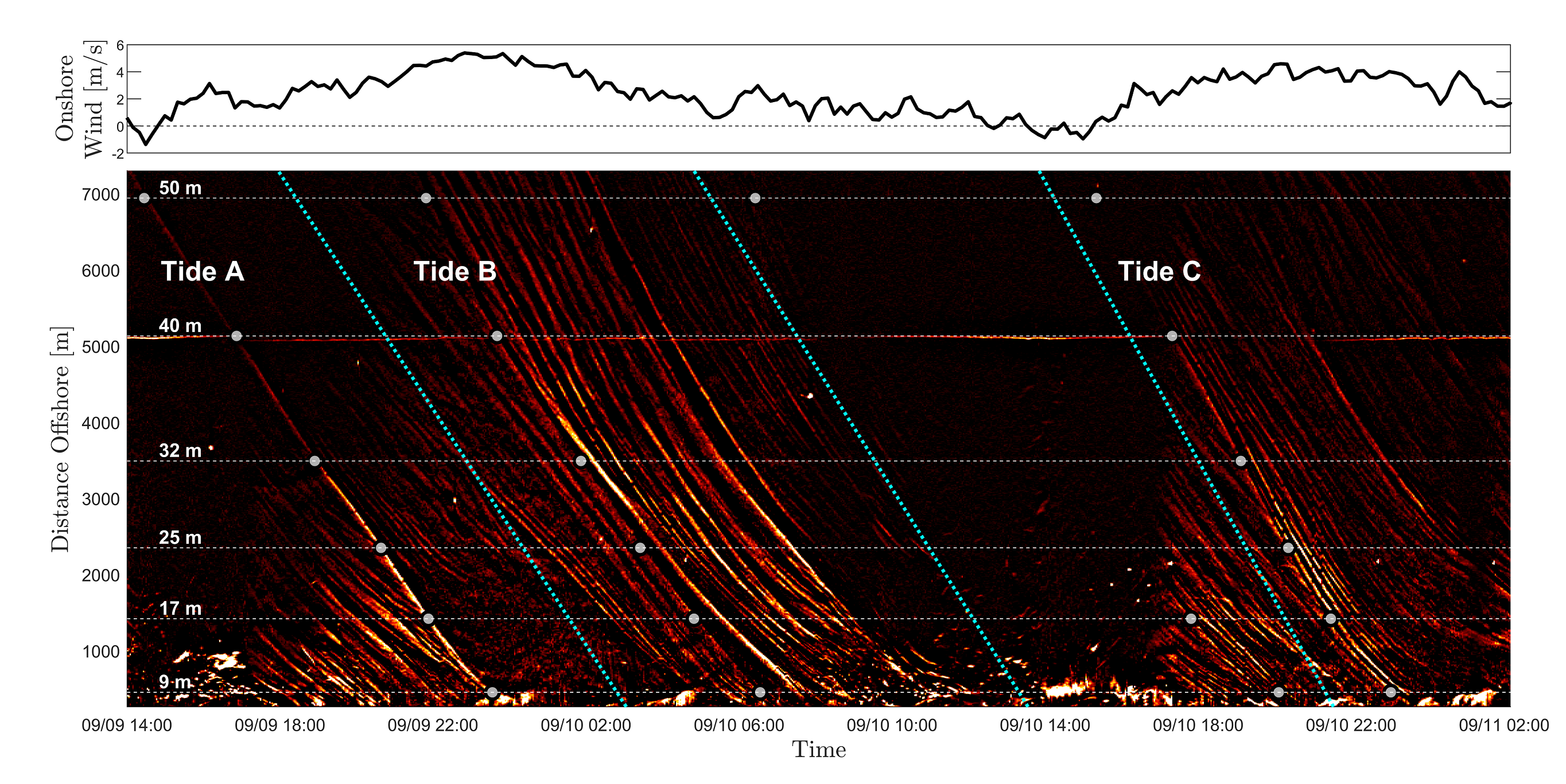}}
\caption{Space-time diagram spanning four internal tides, with the onshore component of wind speed plotted above. The teal dashed lines are temporal delineations between the tides. Grey dots are bore arrivals in the mooring data.}
\label{fig: big timestack}
\end{figure}

We chose three nearly-consecutive internal tides for a detailed analysis of the combined radar and mooring observations due to their strong presence in the radar imagery as well as their distinctness from one another in wave packet compositions. A space-time diagram for the full duration of the three tides is shown in Figure \ref{fig: big timestack}. They occurred between September 9-11, where tides A and B are internal tides spaced by 6 hours, and tide C is an internal tide 12 hours after tide B (skipping one tide). Temporal delineations between the three tides are plotted as diagonal cyan lines; these are chosen qualitatively only to help the reader distinguish between the three tides. Tides A, B, and C were well-imaged due to ample onshore wind (2-4 $ms^{-1}$), as shown in Figure \ref{fig: big timestack}. A wind relaxation event following Tide B prevented radar imaging of the internal tide between B and C, before the wind picked back up again around the timing of Tide C.  The onshore wind component increased on Sep 10 around 16:00, which can be seen in both the wind time series and the space-time diagram, as the intensity increases along the full offshore distance. Thus a short later portion of the internal tide between B and C can be seen. 

 The radar signature of mooring OC40S is visible in the space-time diagram as a horizontal streak roughly 5000 $m$ offshore, demonstrating that the extracted radar transect used to create the space-time diagram overlaps OC40S; the other mooring locations are within 100 $m$ of the transect. For each of the four tides spanning the time period shown in the space-time diagram, the lead bore detected in the mooring observations \citep{mcsweeney2020observations} is back plotted onto the space-time diagram. It is evident that the mooring-detected arrivals well-align with the bright waves in the space-time diagram, with some deviations to be discussed herein. 

\begin{figure}[h]
\centerline{\includegraphics[width=42pc]{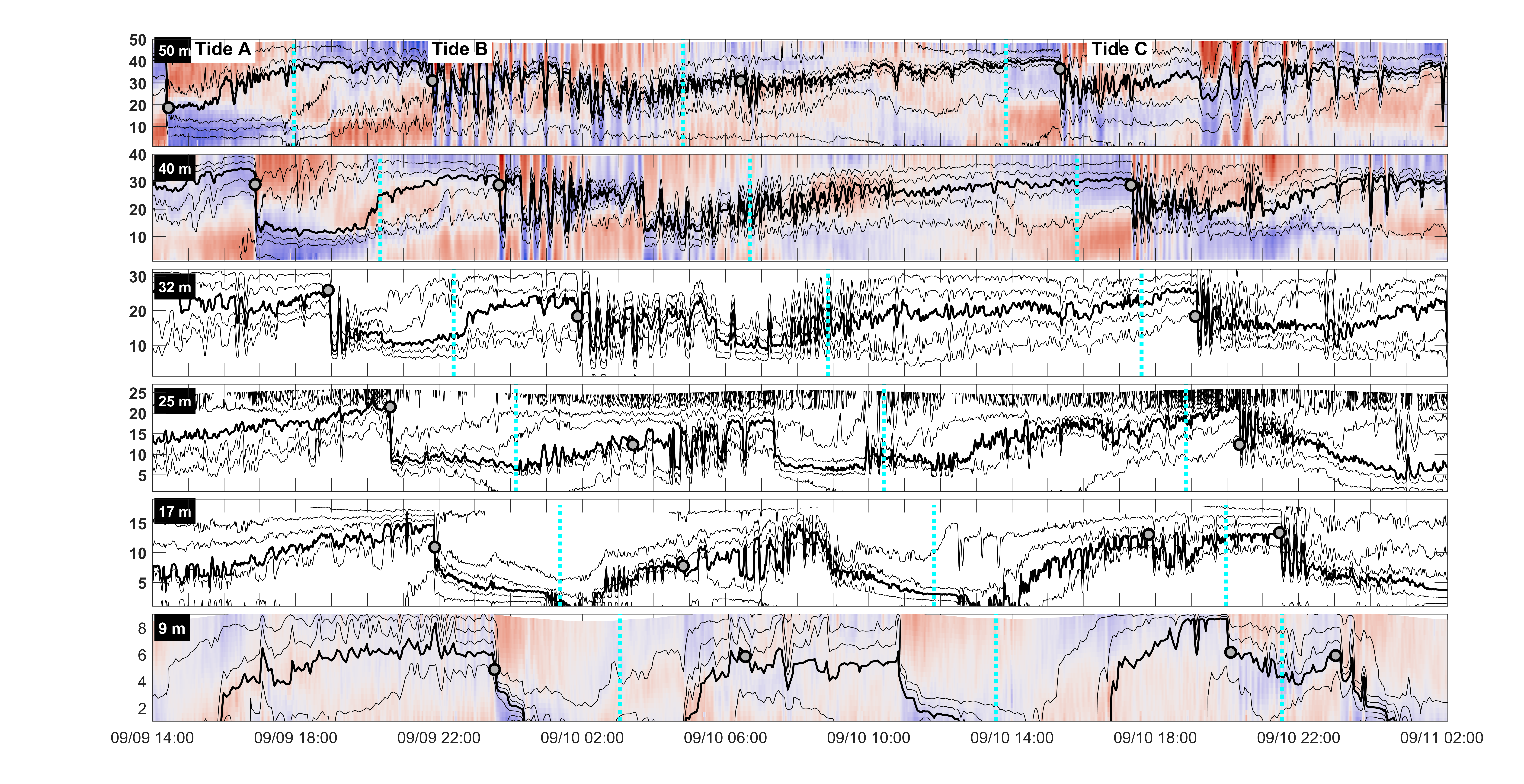}}
\caption{Mooring observations at OC50, OC40S, OC32S, OC25SB, OC17S, and STR3B (refer to Figure \ref{fig: region} for locations). Eastward velocity is plotted in blue and Westward velocity is plotted in red. Temperature is contoured at 1$^{\circ}$ intervals with the 15$^{\circ}$ contour plotted in bold.}
\label{fig: big mooring data}
\end{figure}

Mooring data between 50 $m$ and 9 $m$ water depth for these three tides is shown in Figure \ref{fig: big mooring data}, with OC40S thru STR3B moorings spatially overlapping the transect used for the space-time diagram. Eastward velocity is plotted for moorings OC50, OC40S, and STR3B, with positive values towards the East (roughly onshore); velocity was not recorded for OC32S, OC25SB, or OC17S $m$. Temperature is contoured for each mooring at 1$^{\circ}$ intervals with the 15$^{\circ}$ contour plotted in bold. Tides A-C are delineated using cyan lines, which correspond to the times and locations of the cyan lines in Figure \ref{fig: big timestack}. The arrival of the lead bores for each tide in the mooring data are also plotted as grey dots in Figure \ref{fig: big mooring data}. 

Figure \ref{fig: big timestack} indicates that Tides A-C are visibly different from one another. Tide A is characterized by a bright and prominent lead bore, with some leading waves closer to shore but no trailing waves. Using the radar data alone, it would be relatively straightforward to identify this bore. Tides B and C, however, contain many waves within the packet. Initial comparisons between the tides indicate that the packet of waves in Tide C is both more tightly spaced, as well as faster (noticeably steeper slope) than Tide B. Within all three packets, it is evident that the space-time trajectory of each wave changes with decreasing distance to shore. While Tides A and C contain consistently imaged features from offshore to onshore, Tide B is visibly more complicated, with many of the linear features broken into shorter segments throughout the cross-shore. In the proceeding sections, we will discuss inter-packet speed variability, spatial complexity, and polarity reversal using a close analysis of these three tides.

\subsection{Event-based characteristics of internal tides}

 Tides A and C demonstrate vastly different intra-tidal, or intra-packet wave characteristics. In Tide A, a very large bore behaves nearly independently, with some short-lived trailing and leading waves. Tide C, on the other hand, holds a consistent shape of rank-ordered waves with the lead bore the largest followed by a series of waves of decreasing amplitude. Here we examine each tide in more detail through coupled radar and mooring analysis. 

\begin{figure}[h]
\centerline{\includegraphics[width=40pc]{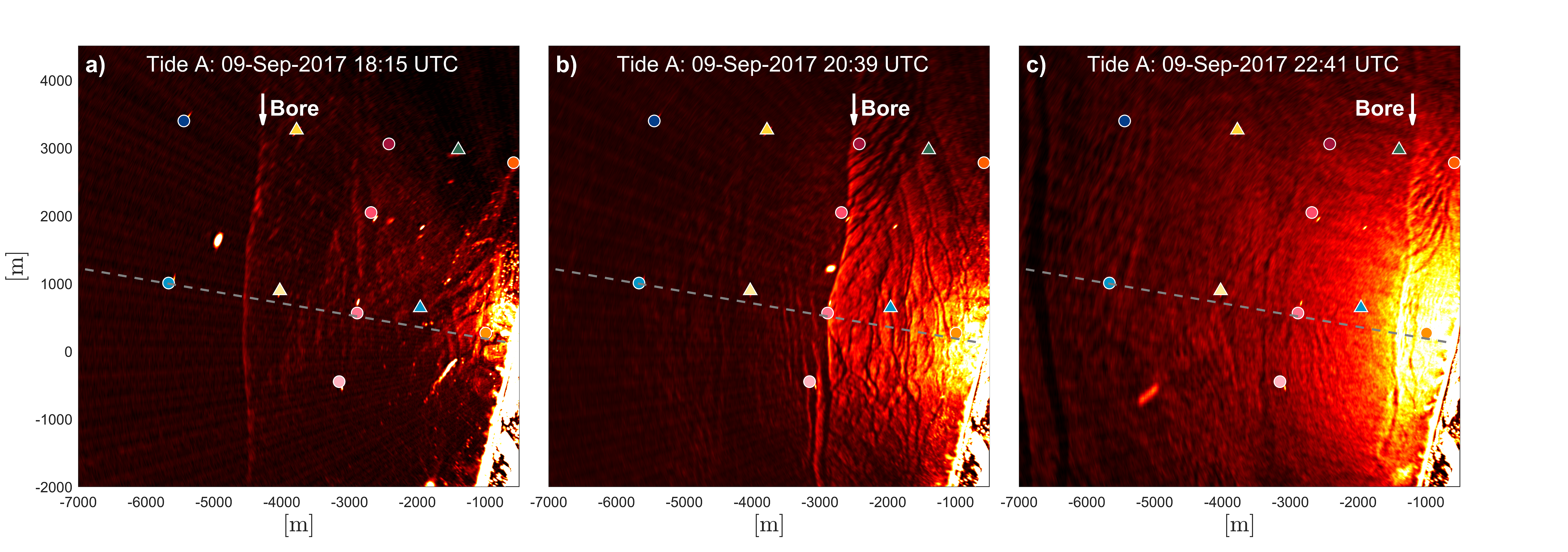}}
\caption{Three frames showing the propagation of Tide A in the radar imagery. The colored symbols are the locations of moorings in the radar footprint (refer to Figure \ref{fig: region}). The dashed grey line is the transect from which the space-time diagram is extracted. The location of the bore is indicated. In frame a) the bore is between OC40S and OC32S. In frame b) the bore is arriving at OC25SB. In frame c) the bore is between OC17S and STR3B.}
\label{fig: tide A three radar figures}
\end{figure}

 As seen in the space-time radar imagery in Figure \ref{fig: tide A three radar figures}, Tide A contains a prominent bright bore that propagates to shore. The lead bore was easily traced using the edge detection routine from 7000m ($>$40m depth) to shore, indicated by a red line. Two trailing waves were additionally identified in the space-time diagram, indicated by blue and green dots at the locations they cross the 32m and 25m moorings. However, approximately 100m after crossing the 25m mooring location, the signatures of these waves appear to slow (indicated by their shallowing slopes) and disappear. The space-time diagram also indicates many higher frequency waves before the main bore, which are traveling slower than the main bore, appear to merge into one another, and are eventually overtaken by the main bore of Tide A. If we turn to the mooring observations, we can see that Tide A contains a large, rarefied lead bore. The mooring-identified bore is shown as a grey dot, and the time-stack identified bore is shown as a red arrow, indicating good agreement between the tracking in each method. The two trailing waves identified in the space-time diagram are plotted on the mooring data as blue and green dots. These two waves are seen at the 32m mooring as waves of elevation trailing the lead bore of depression, and have dramatically shrunk in size by the 25m mooring to the point of barely being detectable in the temperature contours. The leading waves seen in the space-time diagram are also very small amplitude in the mooring data, and are not distinguishable between moorings. Therefore, the behavior of the lead bore overtaking the slower-traveling but leading higher frequency waves can only be detected in the space-time diagram. 

To gain a better picture of the leading and trailing waves surrounding the large bore of depression that characterizes Tide A, we can turn to three radar images that show the bore as it moves to shore. Figure \ref{fig: tide A three radar figures} shows that the bore is uniform in the alongshore for its complete propagation to shore. However, the leading and trailing waves are not. \ref{fig: tide A three radar figures}b shows the lead bore as it crosses the 25m mooring, the leading waves are scalloped in shape. Additionally, these leading waves have variable orientations in the alongshore, and appear to have a much stronger southward orientation on the northern portion of the bore (around OC25N). Interestingly, the trailing waves are much more prominent toward the southern half of the region, and consistent with the space-time diagram and mooring analysis, are only visible between the 32m and 25m moorings. Both the scalloped structures preceding the lead bore, and the minimal presence of trailing waves in this particular packet, are unique characteristics of Tide A.

\begin{figure}[h]
\centerline{\includegraphics[width=40pc]{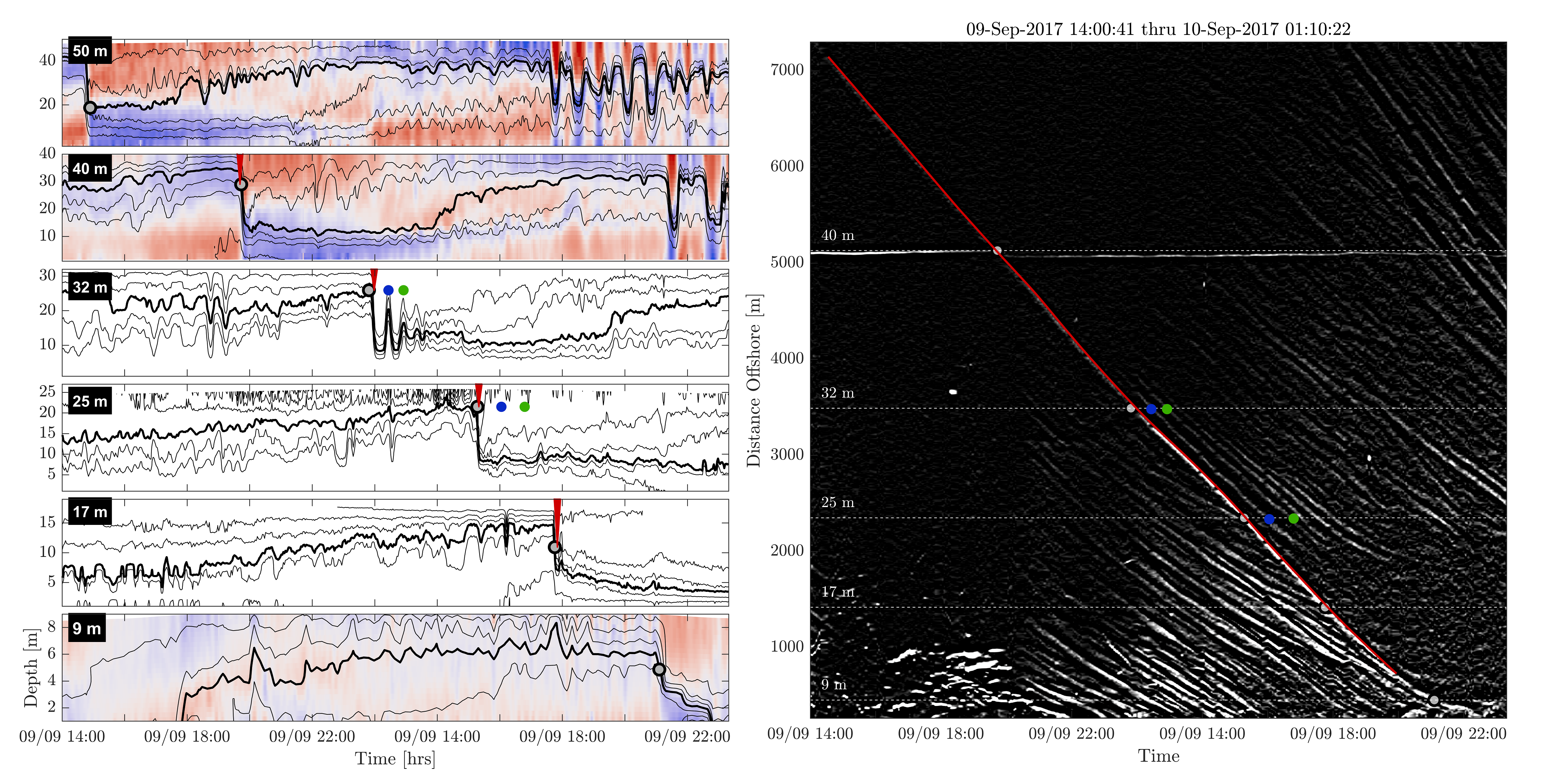}}
\caption{Mooring observations (left) and space-time diagram (right) plotted for the duration of Tide A.}
\label{fig: tide A moor and timestack}
\end{figure}

\begin{figure}[h]
\centerline{\includegraphics[width=40pc]{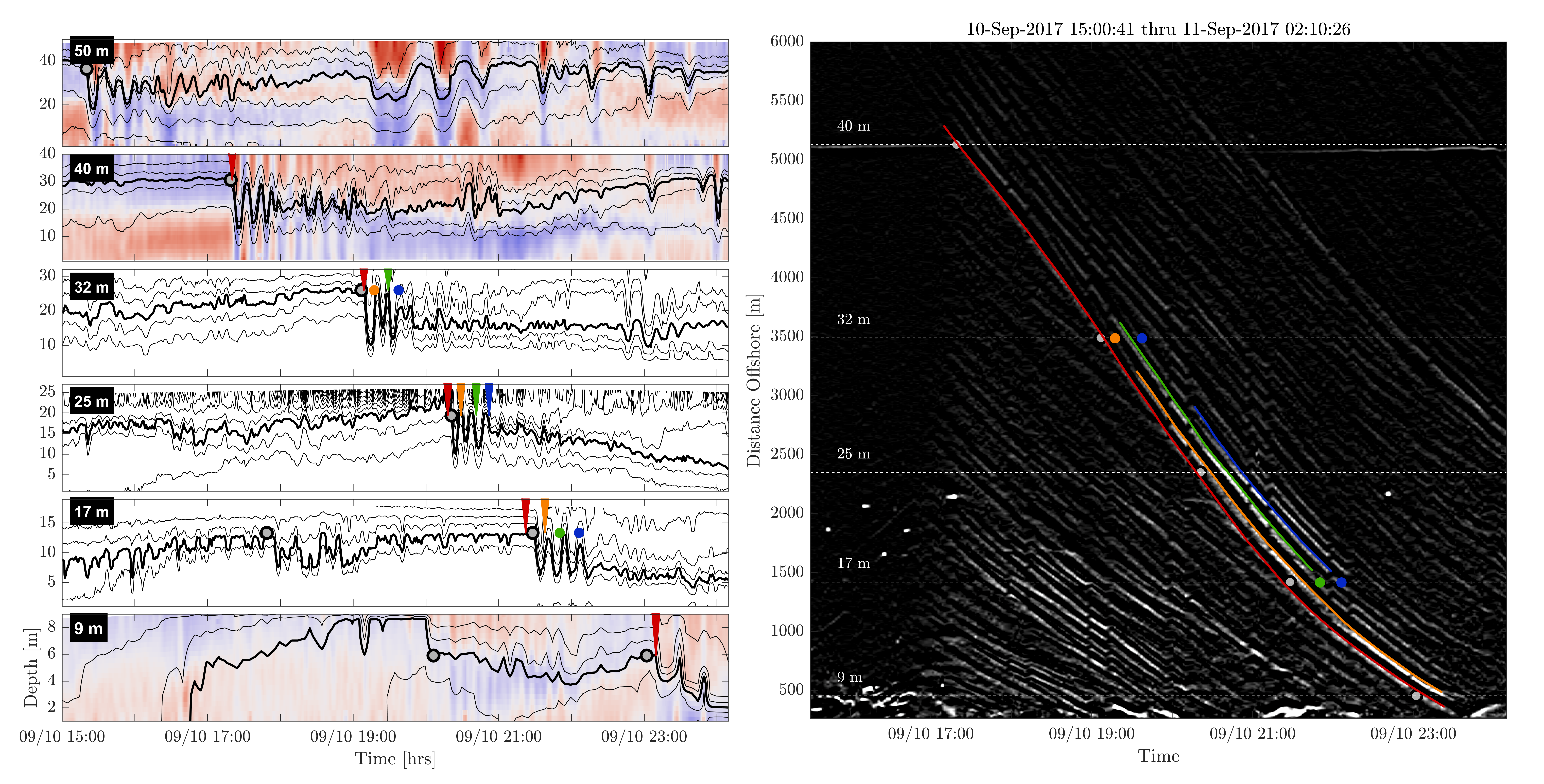}}
\caption{Mooring observations (left) and space-time diagram (right) plotted for the duration of Tide C.}
\label{fig: tide C moor and timestack}
\end{figure}

\begin{figure}[h]
\centerline{\includegraphics[width=40pc]{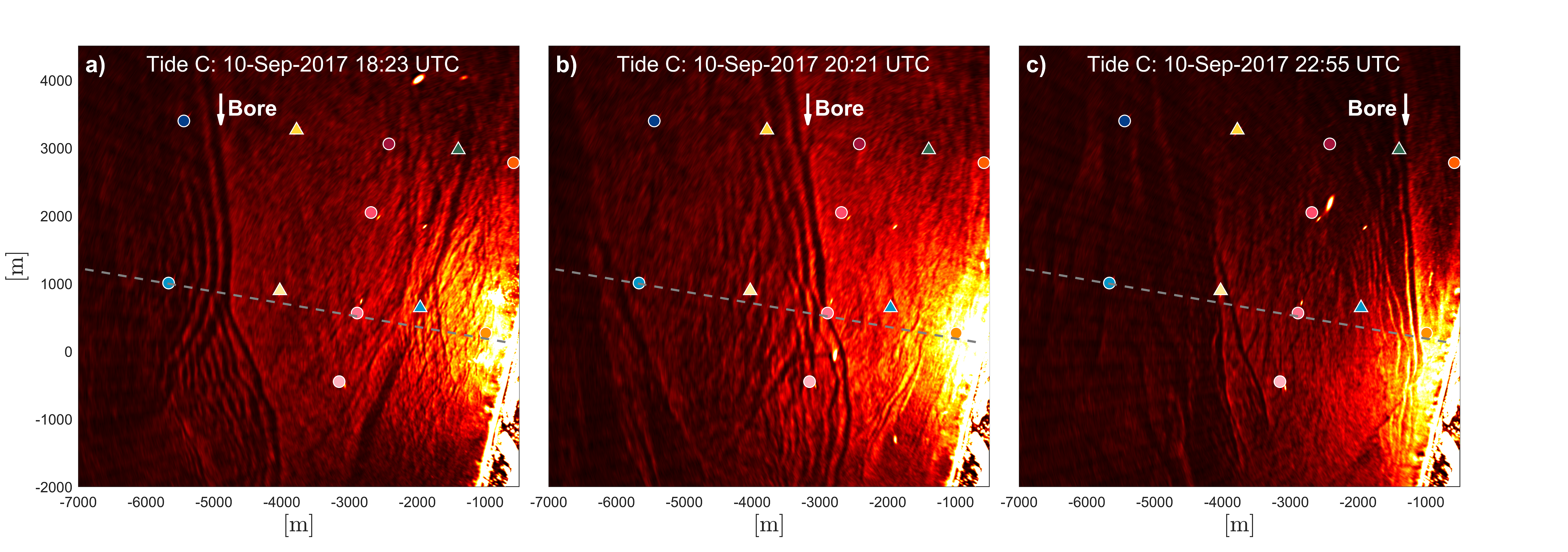}}
\caption{Three frames showing the propagation of Tide C in the radar imagery. The cyan dots are the locations of moorings in the footprint (40-9 $m$ depth). The dashed grey line is the transect from which the space-time diagram is extracted. In frame a the bore is between OC40S and OC32S. In frame b the bore is arriving at OC25SB. In frame c the bore is about to arrive at STR3B.}
\label{fig: tide C three radar figures}
\end{figure}

Tide C, on the other hand, is characterized by a lead bore followed by a clean packet of trailing waves, which are visible in both the space-time diagram and mooring observations (Figure \ref{fig: tide C moor and timestack}). The mooring observations during Tide C, as seen in Figure \ref{fig: tide C moor and timestack}b, reveal a packet of solitary waves of depression at OC50 that maintains its shape as it transits to OC17S. The stratification before the arrival of the bore remains relatively consistent, with the pycnocline above the mid water depth. Similarly, in the space-time diagram (Figure \ref{fig: tide C moor and timestack}b), the packet is well-imaged and shows consistent linear features from 40 $m$ water depth to shore. In addition to the lead bore, the next three waves of the packet have been traced. The intercepts of these traces with the mooring locations are plotted using the same colored triangles on the mooring data in Figure \ref{fig: tide C moor and timestack}a. By looking at Tide C in the radar imagery in Figure \ref{fig: tide C three radar figures}, it is apparent that the waves are less spatially uniform in the alongshore direction, distinctly different from Tide A. In Figure \ref{fig: tide C three radar figures}a, it appears that there are potentially two packets of internal waves that merge just south of the of the transect used for the space-time diagram (dashed grey line).

\begin{figure}[h]
\centerline{\includegraphics[width=40pc]{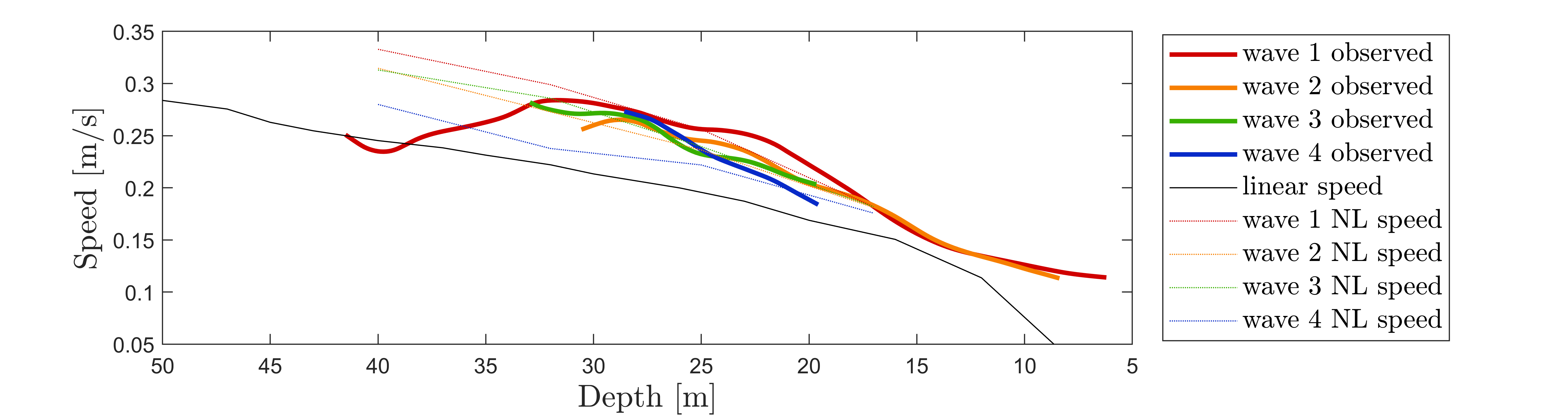}}
\caption{Observed, linear, and nonlinear speeds during Tide C. Solid colored lines represent cross-shore speeds estimated during Tide C from the radar space-time diagram, for the first four waves of the packet. Black solid line represents the linear theory estimated using a 6-hour sorted stratification product. Dashed lines represent the linear speed plus a KdV nonlinear correction term using the amplitude of the first four waves of the packet.}
\label{fig: tide C speeds}
\end{figure}

The continuous cross-shore wave imaging also allows us to estimate the speeds of the waves within a wave packet - something that would be much more difficult using the mooring data alone. Figure \ref{fig: tide C speeds} shows wave speed as a function of cross-shore distance. The color of the speed profiles in Figure \ref{fig: tide C speeds} corresponds to the traced waves in Figure \ref{fig: tide C moor and timestack}b. The black line represents the linear speed estimated from the sorted stratification for this time period. It is important to note that the sorted stratification represents a background, 6 hour moving average product, and thus cannot reveal linear speed estimations on a wave-by-wave basis. This is because the sorted product represents the wave guide that the packet of waves enters into, and not an instantaneous speed prediction for each wave. In order to use the linear speed estimation on a wave-by-wave basis, one would need to solve Equation \ref{eqn: linear speed} at instantaneous density profiles immediately before each wave passes through that stratification. Given that Equation \ref{eqn: linear speed} yields a maximum speed value when the pycnocline is in the middle of the water column, the passing of the first wave will depress the water column to closer to mid water depth, which would solve for a faster second wave than first wave. 

However, nonlinear theory indicates speed should be amplitude dependent (Equation \ref{eqn: speed}), and thus a rank-ordered packet should have decreasing wave speeds on a wave-by-wave basis. In \citet{liu2014tracking}, large amplitude waves O(100m) are observed in SAR satellite data, and nonlinear speed values are estimated using a correction term is applied to linear theory (Equation \ref{eqn: linear speed}) resulting in Equation \ref{eqn: speed}. In \citet{liu2014tracking}, subsurface observations were not available therefore amplitude ($\zeta$) and $\alpha$ values were guessed ($\zeta$ ranging from 120 m to 20m and $\alpha$ of 0.02). The nonlinear correction term indicates that nonlinear speed values will scale with wave amplitude. For Tide C, the observed lead bore speeds are shown in Figure \ref{fig: tide C speeds} as solid lines, with colors corresponding to the identified waves in Figure \ref{fig: tide C moor and timestack}). Indeed, the lead bore is the fastest wave among the packet, with the trailing, smaller waves slightly slower. Here we estimate a nonlinear speed by adding a first-order KdV correction to the linear speed:

\begin{equation}
    c = c_0 + \frac{\alpha \zeta}{3}
    \label{eqn: speed}
\end{equation}
\noindent where $\zeta$ is the internal wave amplitude and $\alpha$ is the nonlinearity coefficient of the KdV equation:
\begin{equation}
    \alpha = \frac{3c_0}{2}\frac{\int_{-H}^{0} (\frac{d\Phi}{dz})^3 dz}{\int_{-H}^{0} (\frac{d\Phi}{dz})^2 dz}
    \label{eqn: alpha}
\end{equation}

The theoretical nonlinear speeds are computed in the cross-shore for the first four waves of the internal wave packet by using $\alpha$ computed at OC40S, OC32S, OC25SB, and OC17S. We manually estimated the amplitude of these waves by looking at each wave in the mooring salinity data. Amplitude values at the 40m mooring decreased from 18m down to 7m from first wave to last wave in the packet. The overall amplitudes of the packets decreased as the packet traveled to shore, resulting in an amplitude range of 7m to 5m at the 17m mooring. An $\alpha$ value of 0.015 $s^{-2}$ was computed using the sorted stratification product, and was found to be roughly consistent for each of the mooring locations. The nonlinear speeds are plotted in Figure \ref{fig: tide C speeds} as dashed lines, where the estimated nonlinear corrections are added to the linear speed (black line). The magnitude of the nonlinear speeds is more comparable to the magnitude of the radar-estimated speeds than the linear speeds, which indicates nonlinearity is a better representation in this case than linear theory. Amplitude-dependent speed decrease between the first wave and last wave is observed in the theoretical nonlinear speeds. While the lead bore is observed to be the fastest wave, the observed speed variation between waves among the wave packet is less distinguishable, and such variation likely falls within the margin of error for observed speed estimation. 

\subsection{Polarity reversal}
\begin{figure}[h]
\centerline{\includegraphics[width=40pc]{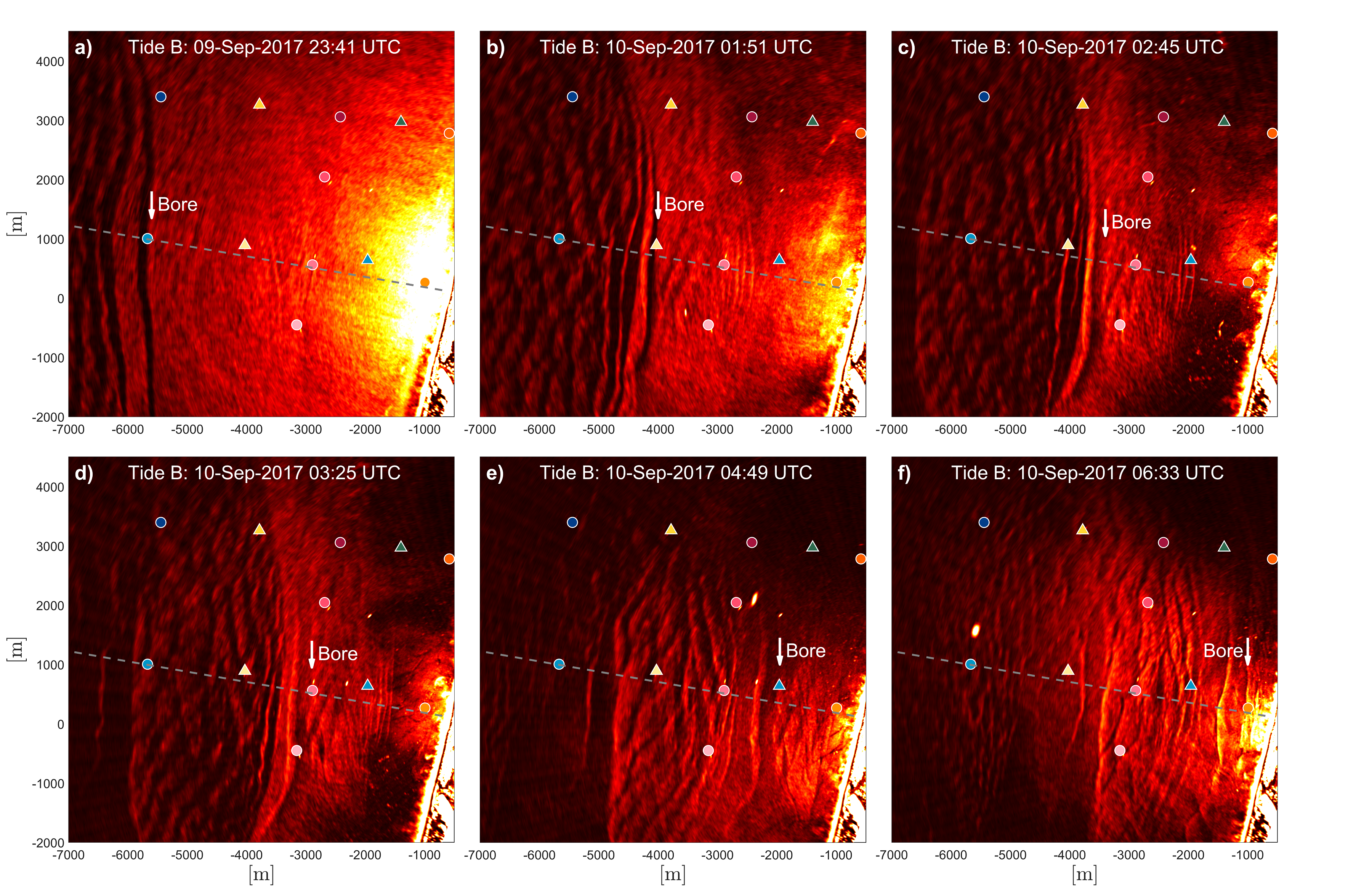}}
\caption{Six frames showing the propagation of Tide B in the radar imagery. The colored symbols are the locations of moorings in the radar footprint (refer to Figure \ref{fig: region}). The dashed grey line is the transect from which the space-time diagram is extracted. The location of the bore is indicated.}
\label{fig: tide B three radar figures}
\end{figure}

\begin{figure}[h]
\centerline{\includegraphics[width=40pc]{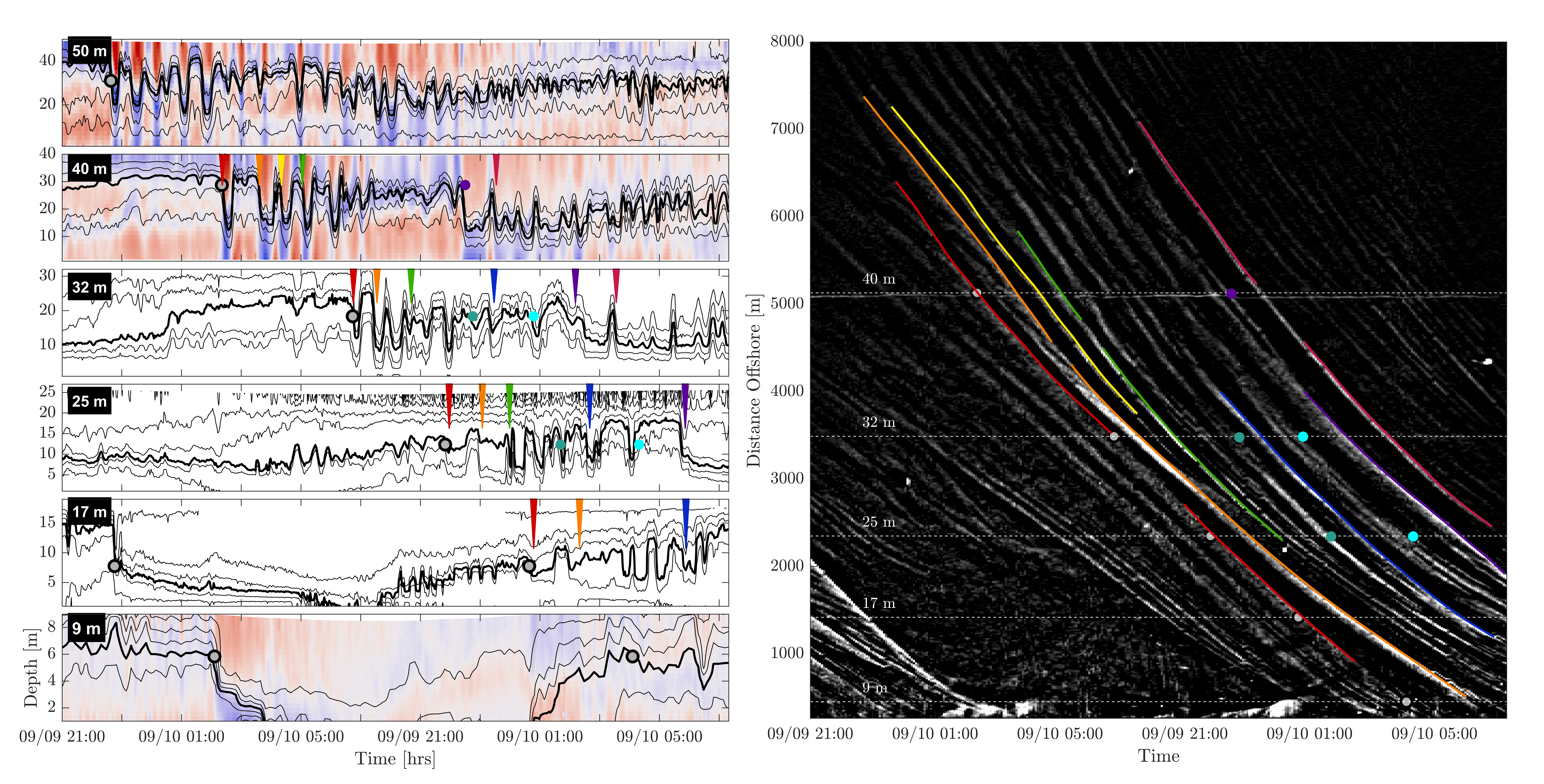}}
\caption{Mooring observations (left) and space-time diagram (right) plotted for the duration of Tide B.}
\label{fig: tide B moor and timestack}
\end{figure}

Figure \ref{fig: tide B three radar figures} shows six snapshots of the propagation of Tide B to shore. Through use of the combined radar and in situ dataset, we have identified an instance of potential polarity reversal during this tide. Tide B begins at OC50 as a packet of internal waves of depression which can be seen in Figure \ref{fig: tide B moor and timestack}. This is to be expected, as the pycnocline has restored to its position higher in the water column (shallower than the critical depth) following tide A. While the radar imaging does not reveal NLIWs at OC50 due to lower signal at far ranges during this time period, it is apparent in the moorings that the NLIWs in the packet have relatively constant temporal spacing throughout the packet (Figure \ref{fig: tide B moor and timestack}a). Between OC50 and OC40S, however, the temporal gap between the lead bore and second wave widens in the mooring data. In the OC40S mooring data, the gap between the first (red) and second (orange) wave is larger than the gap between the first and second waves in OC50, and also wider than the rest of the waves in the packet at OC40S. This is additionally apparent in the radar imagery, as seen in Figure \ref{fig: tide B three radar figures}a. At this time stamp, the packet of waves are arriving at OC40S. There is considerably more spacing between the lead bore and the rest of the packet of NLIWs. Additionally, the offshore radar signature of the wave packet is a series of wide, dark bands with the bright band between the waves the same intensity as the background intensity. The packet is imaged almost entirely via dark signatures evident of divergence, with little to no evidence in the radar imagery of increased backscatter above the background backscatter. This imaging mechanism is unique to this tide when compared to Tides A and C. 

At OC32s, the structure of the first two waves (red and orange) is comparable to OC40S, although the gap between them has shortened. This is evident in both the mooring observations, as well as in the space-time diagram (Figure \ref{fig: tide B moor and timestack}) as the red and orange traces veer closer together near the 32 $m$ depth contour. In looking at the mooring data, the rest of the packet looks somewhat different. At OC40S the packet of waves of depression each restored to a relatively consistent pycnocline depth ($\sim$30 $m$). However, at OC32S, the waves trailing the second (orange) wave do not restore to a comparable depth as the red and orange waves, rather they only restore to $\sim$20 $m$. This is a potential hinting of polarity reversal. The OC25S mooring observations reveal a potential initiation of polarity reversal among the red and orange waves. The pycnocline leading up to the red wave has now depressed to $\sim$15 $m$. The amplitude of the red wave is now quite small, roughly 5 $m$ compared to 10-15 $m$ at OC50 thru OC32S. Between the red and orange wave, there is an evident wave of elevation emerging from the depressed pycnocline. This wave of elevation between the red and orange waves becomes even more apparent at OC17S.  

Tide B additionally illustrates the strong benefits of combining remote sensing observations with in situ observations for tracking of internal bores. In the radar images (Figure \ref{fig: tide B three radar figures}) and space-time diagram (Figure \ref{fig: tide B moor and timestack}), the lead bore is straightforward to identify from 40 to 32 $m$ depths, as it is the leading wave and the brightest feature. However, between 32 $m$ and 25 $m$ (Figure \ref{fig: tide B three radar figures} panel c) the signature behind the lead bore becomes the brightest feature in the radar. This is actually the restored pycnocline between the first (red) and second (orange) waves, which we hypothesize becomes a wave of evolution. In the radar imagery alone (Figure \ref{fig: tide B three radar figures}) the lead bore would not be correctly identified as the imaging changes as the waves propagates to shore. This has implications in the ability to estimate speeds from the radar imagery alone, because the identification would jump between waves causing an artificial jump in speed. In the space-time diagram (Figure \ref{fig: tide B moor and timestack}), the lead and second bores are more visible as distinct features than in the radar images (Figure \ref{fig: tide B three radar figures}), due to the intensity normalization and the ability to track a single feature over time. However, without comparison to the mooring observations, the identified bore would likely be the second wave of the packet, as this is a brighter feature than the first wave.

\subsection{Wave reflection from shore}

\begin{figure}[h]
\centerline{\includegraphics[width=40pc]{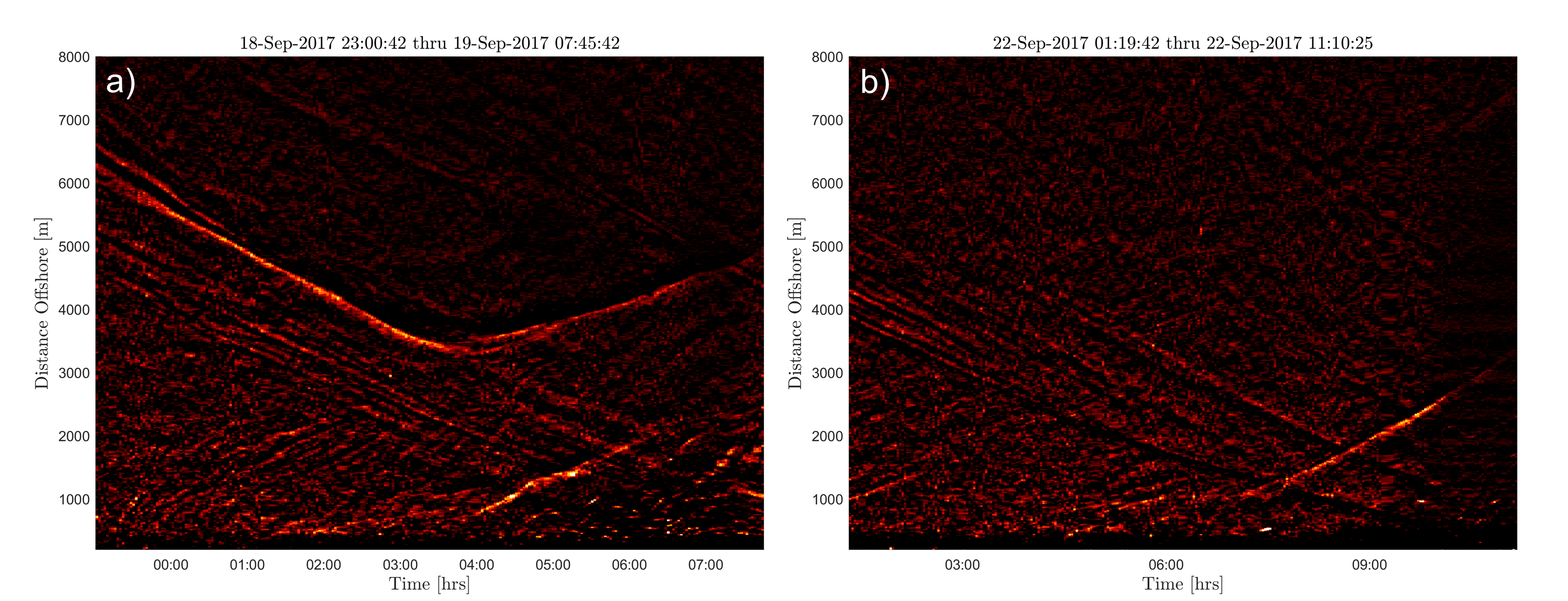}}
\caption{Space-time diagrams from two different time periods showing examples of offshore-propagating features indicative of reflected internal waves.}
\label{fig: reflected}
\end{figure}

Throughout the two month experiment, space-time diagrams were made every 6 hours intended to capture each internal tide. Through visualization of these space-time diagrams we noticed several tides that contained an offshore-propagating bright feature, indicative of potential internal wave reflection. Two representative space-time diagrams are shown in Figure \ref{fig: reflected}. In the left diagram, a bright linear feature propagates to shore with a sharp turnaround at 3500m offshore (32m depth). In this diagram it is quite evident that the off-shore propagating wave is connected to the incoming bore, indicating likely imaging of a reflected internal wave. On the right panel, however, there is an offshore-propagating feature that begins farther inshore at 1000m meters from shore, or 12m water depth. While this offshore propagating feature appears to occur at a time period when a packet of waves is coming to shore, it is not possible to link this feature to potential reflection of a specific incoming internal wave. A similar front is observed inshore at 1000m in the left diagram time as well. While these potential reflected internal waves are not investigated further in this study, we have included them as an indication of the capabilities of this sensing method, as well as motivation for future work. 

\section{Discussion}
The presented speed estimation technique using X-Band radar proves comparable to speed estimation from moorings, while providing higher cross-shore resolution of the shoaling speed profile. Overall cross-shore trends in internal tide bore speed are consistent with the findings of \cite{mcsweeney2020observations}, thereby demonstrating the utility of an X-Band radar as a tool for long-term bore speed estimation. However, interestingly, the lead bore as seen via in situ observations is not consistently the brightest wave in the X-band radar observations. Specifically, around 40 $m$ water depth in Tide B, the in situ observed lead wave is evident as the brightest and leading wave in the X-band radar observations. However, around 32 $m$ depth, the second wave becomes significantly brighter in the X-band radar. It can be seen in the in situ observations that the first wave, between 32 and 25 $m$ depth has transformed from a wave of depression to a wave of elevation, and in the process become a smaller amplitude wave. We hypothesize that this amplitude reduction during the polarity reversal process is leading to the reduction in brightness of the first wave in the radar. With the radar alone, the brighter, second wave is likely to be identified as the leading edge of the wave packet in Tide B, thus missing the first wave observed in the in situ observations. This highlights the importance of the cross-sensor comparison. 

Tides A and C contain solitary waves of depression. While Tide A contains, mostly, a single bore of depression, Tide C is a packet of waves of depression. We show that the first four waves are rank ordered in amplitude, with the lead bore exhibiting the fastest speed. Using a KdV nonlinear correction to the theoretical linear speeds, the nonlinear speed reasonably well predicts the speed of this wave packet. Additionally, while Tide B does not show rank ordered waves in terms of speed, it is evident in the space-time diagram and mooring observations that the wavelength of the first waves is longer than the rest of the packet. Several studies have looked at inter-packet dynamics. In \cite{fu1982seasat}, Seasat SAR images show that there is variability in wavelength within a group of waves in deep water. This study hypothesized that the waves were rank ordered in amplitude, but was not able to confirm without mooring observations. Additionally, \cite{sutherland2001finite} studies wave packet dispersion. Using fully nonlinear finite amplitude simulations, they predict that for small-amplitude waves, the amplitude within a wave packet decreases over time. This is consistent with the findings in Tide C. It is of note that the phase speed of the waves in Tide C are the fastest that were detected using the radar throughout the experiment; however, at this point it is unclear why. For packets with larger amplitude waves, \cite{sutherland2001finite} predict that the amplitude will increase initially, but then decreases as the packet subdivide, which is a possible explanation for the increased spacing that occurs during Tide B. 

Tide B additionally demonstrates a potential polarity reversal from waves of depression to one or multiple waves of elevation. To corroborate the likelihood of this happening, we compute the parameter $\alpha$ (Equation \ref{eqn: alpha}) over depth for each of the three tides. To do this, density measurements at OC50, OC40S, OC32S, OC25S, OC17S, and STR3B are averaged 1 hour before the arrival of the bore at each mooring. A 1 hour window was chosen as the representative wave guide into which the packet of NLIWs was entering. The values of $\alpha$ for Tides A and C fall between 0.01 and 0.02 $s^{-2}$, while values of $\alpha$ are 0.01 $s^{-2}$ at the 40m mooring but dip to 0 $s^{-2}$ at the 30m mooring and -0.01 $s^{-2}$ by the 17m mooring. Positive values of alpha indicate the pycnocline is above mid water-column, and NLIWs should be waves of depression. A sign change in $\alpha$ to negative indicates the pycnocline has dropped below mid water-column, and waves of elevation are likely. In both Tides A and C, values of alpha remain positive through all depths, and, only waves of depression are observed. In Tide B, however, the decrease in $\alpha$ to a negative value supports the observations that initial waves of depression offshore undergo a polarity reversal into waves of elevation beginning around the 25 $m$ depth contour. 

Contrary to the theory of weakly nonlinear amplitude dispersion, which predicts a rank-ordered decrease in wave speed throughout the packet, linear theory predicts an increased wave speed throughout a wave packet due to the individual waves' influence on the stratification. In other words, the waves themselves influence the wave guide for the proceeding waves. According to Equation \ref{eqn: speed}, the stratification that produces the fastest wave speed will be when the pycnocline is at the critical wave depth, or perfectly mid-depth. Typically, the stratification is elevated above the critical depth, or on rarer occasions, depressed below the critical depth, before the arrival of the wave. Therefore, the depression or elevation of the pycnocline due to the presence of the waves to the critical depth will increase the theoretical linear wave speed. 

Last, we discuss the influences that internal tides may have on each other. It is of particular note that at OC50, OC40S, and OC32S, the pycnocline between Tides A and B restores too a depth above mid-water column. However, at OC25S and shoreward, the pycnocline remains depressed below mid water-column between Tides A and B. This is a potential cause of the complicated dynamics of this tide, and highlights the importance of tide-to-tide influences on the stratification. 

\section{Conclusions}
We have analyzed observations of internal waves in a unique dataset collected during the Inner Shelf Dynamics Experiment, September-October 2017. The observations consist of X-Band radar imagery gathered over a 10 $km$ radius of the nearshore inner shelf, as well as stationary mooring observations. Not only is this a novel dataset due to its high spatial and temporal resolution and long deployment duration, it is additionally a novel study due to the relatively shallow water in which the waves are observed. Thus, we have provided an overview of observed transformation dynamics as the waves shoal, beginning in depths $<$40 $m$ and eventually to shore dissipation. The developed detection routine using the X-Band radar uses space-time diagrams along cross-shore transects to detail internal wave speed and angle with resolution of O(5 $m$) and O(2 $sec$). We analyze 7 leading bores (largest wave in internal tide) for an overall cross-shore speed trend, which indicates that the internal bores propagate with steady speed that is slower than linear theory until $\sim$32 $m$ depth, surpasses linear theory, and then decreases at a rate faster than linear theory would predict. In addition to analysis of lead bores, the remote sensing technique allows for in depth analysis of multiple waves within a packet, which is traditionally difficult to analyze over a transformation region from mooring data alone. We have detailed three nearly-consecutive tides using both X-Band radar and mooring observations to reveal evidence of polarity reversal and inter-packet dispersion dynamics. Wave-to-wave variability among packets indicates that the waves are rank ordered, with the lead bore being the fastest wave. Additionally, by looking in detail at three packets, we have identified likely tide-to-tide influences on the stratification that influence consecutive packet behavior. In the instance of observed polarity reversal dynamics, the second waves appears to transform from a wave of depression to a wave of elevation, and computed values of the KdV nonlinearity coefficient $\alpha$ switches signs indicating that a polarity reversal is likely. Using the X-band radar, the ability to track single features consistently through space makes it possible to reveal inter-packet dynamics which poses a challenge to the point observations of moorings alone. However, the presence of in situ observations with spatial and temporal overlap to the radar remains critical for understanding the subsurface shape and behavior of the remotely sensed internal waves.

%

%

\clearpage
\acknowledgments

%
%
\datastatement

%






%



\bibliographystyle{ametsocV6}
\bibliography{references}

\end{document}